\documentclass[useAMS,usenatbib]{mn2e}
\usepackage{graphicx}
\usepackage{rotfloat}

\newcommand{\ms}{$\,$M$_\mathrm{\odot}$}

\newcommand{\be}{\begin{equation}}
\newcommand{\ee}{\end{equation}}
\newcommand{\stars}{{\sc stars}}
\newcommand{\el}[2]{\ensuremath{^{#1}\mathrm{#2}}}

\newcommand{\ag}{\ensuremath{\mathrm{(\alpha,\gamma)}}}
\newcommand{\an}{\ensuremath{\mathrm{(\alpha,n)}}}

\title[Light elements in CEMP stars]{Light element abundances in carbon-enhanced metal-poor stars}
\author[R.~J. Stancliffe]{Richard J. Stancliffe\thanks{E-mail:
Richard.Stancliffe@sci.monash.edu.au}\\
Centre for Stellar and Planetary Astrophysics, Monash University, PO Box 28M, Clayton VIC 3800, Australia \\
}
\begin{document}
\bibliographystyle{mn2e}

\date{Accepted 0000 December 00. Received 0000 December 00; in original form 0000 October 00}

\pagerange{\pageref{firstpage}--\pageref{lastpage}} \pubyear{0000}

\maketitle

\label{firstpage}

\begin{abstract}
We model the evolution of the abundances of light elements in carbon-enhanced metal-poor (CEMP) stars, under the assumption that such stars are formed by mass transfer in a binary system. We have modelled the accretion of material ejected by an asymptotic giant branch star on to the surface of a companion star. We then examine three different scenarios: one in which the material is mixed only by convective processes, one in which thermohaline mixing is present and a third in which both thermohaline mixing and gravitational settling are taken in to account. The results of these runs are compared to light element abundance measurements in CEMP stars (primarily CEMP-$s$ stars, which are rich in $s$-processes elements and likely to have formed by mass transfer from an AGB star), focusing on the elements Li, F, Na and Mg. None of the elements is able to provide a conclusive picture of the extent of mixing of accreted material. We confirm that lithium can only be preserved if little mixing takes place. The bulk of the sodium observations suggest that accreted material is effectively mixed but there are also several highly Na and Mg-rich objects that can only be explained if the accreted material is unmixed. We suggest that the available sodium data may hint that extra mixing is taking place on the giant branch, though we caution that the data is sparse.
\end{abstract}


\section{Introduction}
Carbon-enhanced metal-poor (CEMP) stars are defined as stars with [C/Fe]\footnote{[A/B] = $\log (N_\mathrm{A}/N_\mathrm{B}) - \log (N_\mathrm{A}/N_\mathrm{B})_\odot$}$>$+1.0 \citep{2005ARA&A..43..531B}, with [Fe/H]$\ <-2$ in most cases. These objects appear with increasing frequency at low metallicity \citep{2006ApJ...652L..37L}. The study of CEMP stars is being used to probe conditions in the early universe. For example, CEMP stars have been used to infer the initial mass function in the early Galaxy \citep[e.g.][]{2005ApJ...625..833L}. Chemical abundance studies have revealed that the majority of the CEMP stars are rich in $s$-process elements like barium \citep{2003IAUJD..15E..19A}, forming the so-called CEMP-$s$ group. Recent survey work has detected radial velocity variation in around 68\% of these CEMP-$s$ stars and this is consistent with them all being in binary systems \citep{2005ApJ...625..825L}.

Binary systems provide a natural explanation for these objects, which are of too low a luminosity to have been able to produce their own carbon. The primary\footnote{This is the initially more massive star in the system; the secondary is the initially less massive star.} of the system was an asymptotic giant branch (AGB) star which became carbon-rich through the action of third dredge-up (the deepening of the convective envelope into regions of the star where material has undergone nuclear burning, see e.g. \citealt{1983ARA&A..21..271I}) and transferred material on to the low-mass secondary (most likely via a stellar wind). The primary became a white dwarf and has long since faded from view, with the carbon-rich secondary now being the only visible component of the system.

It has commonly been assumed that accreted material remains unmixed on the surface of the recipient star during the main sequence and only becomes mixed with the stellar interior during first dredge-up. However, \citet{2007A&A...464L..57S} pointed out that this picture neglects thermohaline mixing. Thermohaline mixing is the process that occurs when the mean molecular weight of the stellar gas increases toward the surface. A gas element displaced downwards and compressed will be hotter than its surroundings. It will therefore lose heat, become denser and continue to sink. This leads to mixing on thermal timescales until the molecular weight difference is eliminated \citep{1972ApJ...172..165U,1980A&A....91..175K}.  Accreted material has undergone nuclear processing in the interior of the companion AGB star and hence has a higher mean molecular weight than the pristine material of the companion. The accreted AGB material should therefore become mixed with the pristine material of the secondary because of the action of thermohaline mixing. In their models, \citet{2007A&A...464L..57S} showed that thermohaline mixing could, under optimal circumstances, lead to the accreted material being mixed with nearly 90 per cent of the secondary.

The extent to which accreted material is mixed into the accretor during the main sequence has been questioned. Using a sample of barium-rich CEMP stars (i.e. a set of stars where we expect the AGB mass transfer scenario to hold), \citet{2008ApJ...678.1351A} showed that the distribution of [C/H] values in turn-off stars (i.e. those stars that have reached the end of their main-sequence lives, are still of low-luminosity and have yet to become giants) was different from that in giants suggesting that significant mixing only happened at first dredge-up. A similar point was made by \citet{2008ApJ...679.1541D} using the data of \citet{2006ApJ...652L..37L}. These authors showed that the abundance patterns of turn-off stars and giants were consistent with coming from the same distribution if first dredge-up resulted in the [C/H] value (and also the [N/H] value) being reduced by around 0.4 dex. This drop in both C and N is the result of the accreted material being mixed with pristine stellar material during the deepening of the stellar envelope at first dredge-up. They find that this result is consistent with having an accreted layer of material mixed to an average depth of about 0.2\ms\ (or alternatively having an accreted layer of 0.2\ms\ that remains unmixed until first dredge-up). \citet{2008ApJ...677..556T} also questioned the efficiency of mixing in the binary system CS~22964-161. This system consists of two turn-off stars which are both carbon-rich. The system is also found to be lithium-rich, with $\epsilon(\mathrm{Li})=+2.09$\footnote{$\epsilon(\mathrm{Li})=\log_{10}\left({N_\mathrm{Li}\over N_\mathrm{H}}\right)$+12 where $N$ is the number abundance of the element.}. As lithium is such a fragile element and is easily destroyed at temperatures above around $10^6$\,K, any extensive mixing of accreted material will result in lithium being depleted. \citet{2008ApJ...677..556T} suggest that gravitational settling could in principle inhibit the action of thermohaline mixing as helium diffusing away from the stellar surface would create a stabilising mean molecular weight gradient (a so-called `$\mu$-barrier').

This idea was tested by \citet{2008MNRAS.389.1828S}, who included the physics of both thermohaline mixing and gravitational settling in their models of stars accreting material from a putative AGB donor. Their grid of models over a range of accreted masses  (0.001-0.1\ms) from donors of different initial mass (1-3.5\ms) showed that gravitational settling could only seriously inhibit thermohaline mixing if a small quantity of material had been accreted. \citet{2008MNRAS.389.1828S} compared their predicted surface abundances of carbon and nitrogen to those of observed CEMP stars, concluding that none of their model sets did a good job of reproducing the observed abundances patterns. The aim of this work is to extend the secondary models of SG08 to a greater range of isotopes (rather than just the C and N looked at in that paper) in the hope that these additional isotopes may help to illuminate the processes that happen in the AGB mass transfer formation scenario.

Recently, there has been considerable progress in measuring the abundances of light elements in carbon-enhanced metal-poor stars. \citet{2008ApJ...678.1351A} have provided a sample of over 70 barium-rich CEMP stars that have other abundance determinations, including for sodium and magnesium. The element fluorine has also been measured for the first time in a CEMP star \citep{2007ApJ...667L..81S}. Each element that we have measurements for potentially gives us another way to explore both the nature of the donor AGB stars and what happens to the material that is accreted on to the secondary star. This paper looks at what the light elements in CEMP stars can tell us about these topics. The AGB mass transfer scenario is likely to apply to the majority of CEMP stars (specifically those with $s$-process enrichments which are classified as CEMP-$s$ stars), but not all CEMP stars form this way. In addition, the elements discussed in this paper do not necessarily come from AGB stars (for example, Na and Mg in the Galaxy mostly come from supernovae). The reader should bear these points in mind throughout.

\section{The stellar evolution code}

Calculations in this work have been carried out using a modified version of the \stars\ stellar evolution code originally developed by \citet{1971MNRAS.151..351E} and updated by many authors \citep[e.g.][]{1995MNRAS.274..964P}. The code solves the equations of stellar structure and chemical evolution in a fully simultaneous manner, iterating on all variables at the same time in order to converge a model \citep[see][for a detailed discussion]{2006MNRAS.370.1817S}. The version used here includes the nucleosynthesis routines of \citet{2005MNRAS.360..375S} and \citet{stancliffe05}, which follow the nucleosynthesis of 40 isotopes from D to \el{32}{S} and important iron group elements. The code uses the opacity routines of \citet{2004MNRAS.348..201E}, which employ interpolation in the OPAL tables \citep{1996ApJ...464..943I} and which account for the variation in opacity as the C and O content of the material varies.

To produce our models, we follow the procedure of \citet{2008MNRAS.389.1828S}, hereinafter SG08. We take the average composition of the ejecta (selected abundances are shown in Table~\ref{tab:abundances}) from the 1.5\ms\ model of SG08 and accrete it on to our secondaries at a rate of $10^{-6}$\ms\,yr$^{-1}$. The SG08 AGB models have been evolved with the same code used here \citep[see e.g.][for details]{2004MNRAS.352..984S}, so these models remain fully self-consistent. In each case, we accrete enough material to make the final mass of the star equal to 0.8\ms, which is the appropriate turn-off mass for the halo. The accretion event takes place at a time appropriate to the age at which the primary enters its superwind phase. We use initial secondary masses of 0.7, 0.75, 0.78, 0.79, 0.795, 0.798 and 0.799\ms\ and the metallicity is $Z=10^{-4}$ (corresponding to [Fe/H]$\ =-2.3$). Three model sets have been evolved: a `standard' model which includes only canonical mixing processes (i.e. convection), the second includes thermohaline mixing of the accreted material and the third sequence includes both thermohaline mixing and gravitational settling. Mass loss has not been taken into account in the evolution of these models.

\begin{table*}
\begin{center}
\begin{tabular}{ccccccccccc}
Mass of AGB & \el{1}{H} & \el{12}{C} & \el{14}{N} & \el{7}{Li} & \el{19}{F} & \el{23}{Na} & \el{24}{Mg} & \el{25}{Mg} & \el{26}{Mg} & \el{56}{Fe} \\
donor (\ms) \\
\hline
1.5 & 0.690 & 1.22(-3) & 2.78(-5) & 1.79(-11) & 8.83(-7) & 5.09(-6) & 3.81(-6) & 4.00(-6) & 2.23(-6) & 5.74(-6) \\
2 & 0.678 & 1.62(-2) & 2.61(-5) & 1.18(-11) & 7.95(-7) & 1.66(-5) & 8.56(-6) & 2.45(-5) & 2.17(-5) & 5.61(-6) \\
3.5 & 0.679 & 2.09(-3) & 6.57(-3) & 6.45(-13) & 2.14(-9) & 5.89(-5) & 5.25(-6) & 2.72(-5) & 7.55(-5) & 5.70(-6) \\
Secondary & 0.749 & 1.73(-5) & 5.30(-6) & 4.68(-11) & 2.03(-9) & 1.67(-7) & 2.57(-6) & 3.83(-7) & 3.88(-7) & 5.85(-6)\\
\hline 
\end{tabular}
\end{center}
\caption{The abundances of selected isotopes in the AGB ejecta used in these simulations along with the initial abundances used in the secondary models. The abundances are in the format $n(m) = n\times10^{m}$.}
\label{tab:abundances}
\end{table*}

\section{Results}
One caveat of the models must be borne in mind throughout. The AGB models from which the nucleosynthesis is taken and the detailed models of the secondaries presented here have been calculated using solar-scaled abundances and opacities. The former should only be important for secondary species (i.e. those that derive from metals already present in the star at the time of formation). Most of the species that we are concerned with are primary and have been manufactured from hydrogen and helium within the star. The yields of these primary isotopes should be independent of the choice of initial abundance. As for the opacities, it would be more appropriate to use a set of tables that accounts for the enhancements of the $\alpha$-elements expected in low-metallicity objects. However, an $\alpha$-enhanced set of tables with variable C and O composition as used by the code is not currently available. We would expect that an $\alpha$-enhanced mixture (with increased abundances of \el{16}{O}, \el{20}{Ne} and \el{24}{Mg}) would affect the yields of \el{22}{Ne}, \el{23}{Na} and \el{25,26}{Mg}. As the production pathways for these elements are quite complex and interdependent, it is not immediately obvious how they would change. Computation of such models merits future study and is beyond the scope of this work.

\subsection{Lithium}

\begin{figure}
\includegraphics[width=\columnwidth]{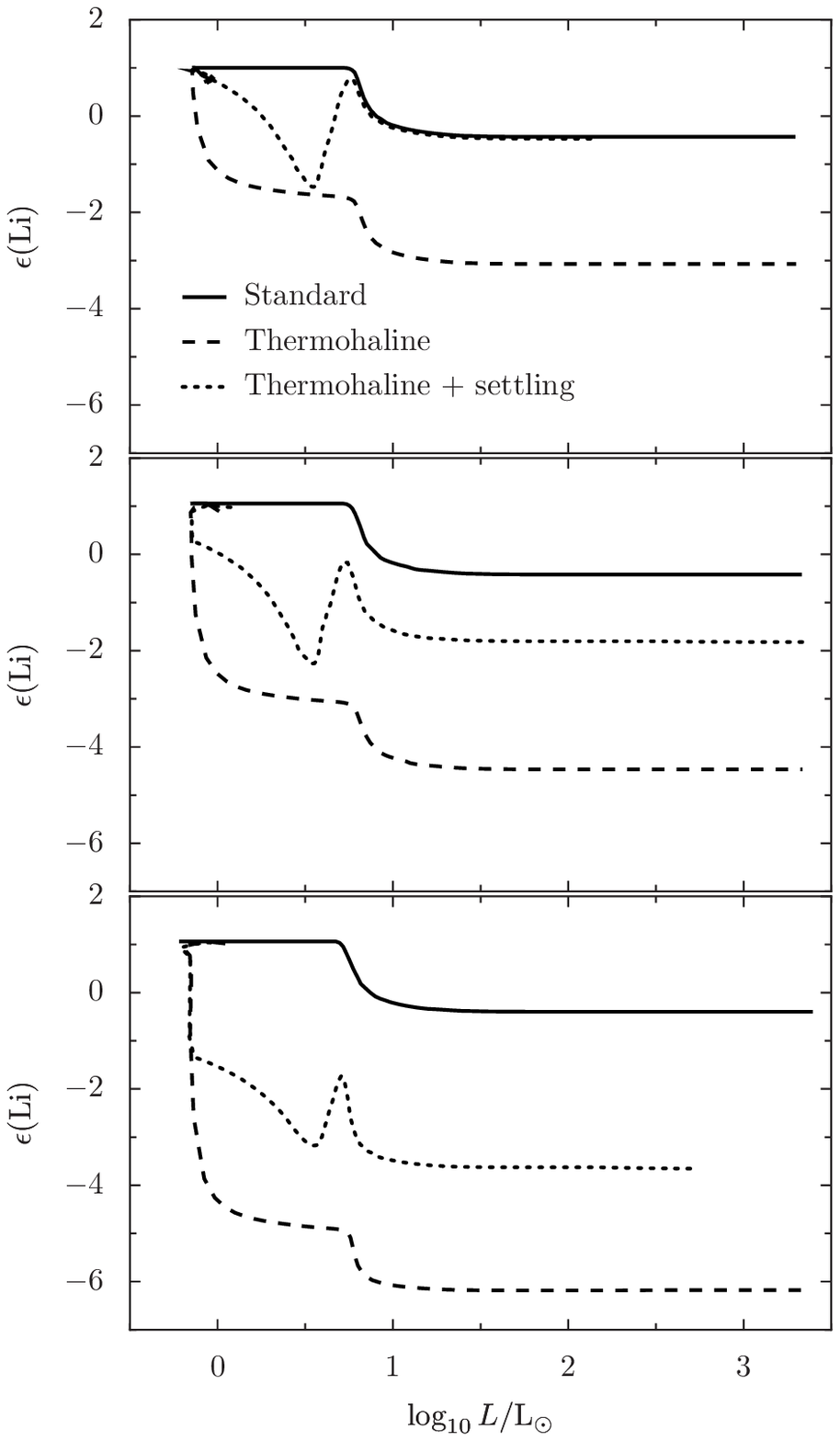}
\caption{The evolution of the lithium abundances as a function of luminosity for each of the model sequences. The standard models are displayed with solid lines and those with only thermohaline mixing are displayed with dashed lines. The models with both thermohaline mixing and gravitational settling are shown by dotted lines. From top to bottom, the panels represent the case of accreting 0.002, 0.01 and 0.1\ms\ of material respectively. In each case, material has been accreted from a 1.5\ms\ companion and the final mass of the star is 0.8\ms.}
\label{fig:LiStack}
\end{figure}

Figure~\ref{fig:LiStack} shows the evolution of the lithium abundance as a function of luminosity for a selection of the models. The standard models show the least depletion of lithium because the accreted material remains in layers that are too cool to destroy this fragile element. The only reduction comes at first dredge-up when the deepening convective envelope mixes the accreted material with the interior of the star. Models including only thermohaline mixing show the greatest degree of depletion, with the reduction being greatest when the most mass is accreted. This is because material is mixed deep into the star, dragging lithium that would normally (in the absence of any mixing) remain near the stellar surface to depths where it can be destroyed. In fact, the presence of Li in CEMP stars could be the best test to determine whether mixing has occurred after the accretion process. A high lithium abundance precludes the occurrence of efficient thermohaline mixing alone (however, see below). The addition of gravitational settling alleviates the destruction of Li somewhat, with a drop in the degree of reduction. In the case that only a small quantity of material is accreted, the abundance can be preserved at close to the accreted value (though gravitational settling causes this value to drop over the main sequence). 

The initial lithium abundance used in both the AGB models and the secondaries are most likely too low. We should expect the initial value to reflect the Spite plateau value of $\epsilon(\mathrm{Li}) \approx2.3$ \citep{1982A&A...115..357S}. However, this may not matter for the AGB models, as lithium should easily be destroyed in the course of a star's evolution. In fact, in the 1.5\ms\ model, the lithium abundances drops by over two orders of magnitude on the first giant branch. For the 1.5\ms\ accreted material, we would not expect the AGB star to have produced any significant amounts of Li during its TP phase (although see \citealt{1997MNRAS.289L..11A} and \citealt{2007A&A...471L..41U} who report the detection of Li in low-mass Galactic AGB stars). Would an enhanced Li abundance in the secondary make a difference? If mixing is extensive, then Li will be depleted and an enhanced `resevoir' in the surface layers will not matter. If the mixing of accreted material is shallow then an enhanced Li abundance in the secondary would lead to a higher abundance after mixing.

Large lithium abundances have been measured in some CEMP stars. For example, CS~22964-161 is a binary CEMP system which shows a lithium abundance of  $\epsilon(\mathrm{Li})=2.09$ \citep{2008ApJ...677..556T}. The AGB models do not have sufficiently high lithium abundance to match this value, even before any mixing of the accreted material takes place (whether this be thermohaline mixing after accretion or convective mixing during first dredge-up). While it is possible for the more massive AGB stars to produce lithium via the Cameron-Fowler mechanism, such a process does not operate in the low-mass AGB stars. We would have to suppose the existence of some additional mixing mechanism, active during the AGB, which would cause circulation of material below the base of the convective envelope down to temperatures where lithium production can take place. Such a supposition may be reasonable in light of the existence of Li-rich low-mass AGB stars in the Galaxy \citep{1997MNRAS.289L..11A, 2007A&A...471L..41U}.

Attention should also be drawn to the Li-rich star HKII~17435-00532. This star has $\epsilon(\mathrm{Li})=2.1$ as reported by \citet{2008ApJ...679.1549R}. The exact evolutionary status of this object is unclear, but it is clear that it has evolved at least as far as the red giant branch -- yet it is still Li-rich! Roederer et al. suggest that it cannot have acquired its Li from a companion: other elements show evidence for a large degree of dilution. This would require the accreted material to have an unreasonably high Li abundance. They discuss potential sources for Li production, but these all occur at late stages in the stars evolution. Unlike HKII~17435-00532, CS~22964-161 is still at a relatively early stage in its evolution and could not have undergone the proposed Li-enrichment events.  We must therefore seek a different explanation for its Li-enrichment.

Recent observations of stars in the metal-poor ([Fe/H]~=~-2) globular cluster NGC 6397 \citep{2007ApJ...671..402K, 2008A&A...490..777L} suggest that models that include  gravitational settling are able to reproduce the abundance trends of certain elements if an additional turbulent mixing process is included \citep{2005ApJ...619..538R}. We have therefore run a model sequence adopting the following prescription \citep[from][]{2005ApJ...619..538R} for such an extra mixing process:
\be
D = 400D_\mathrm{He}(T_0)\left[\rho\over \rho(T_0)\right]^{-3}
\ee
where $D$ is the turbulent diffusion coefficient, $D_\mathrm{He}(T_0)$ is the atomic diffusion coefficient for helium at a temperature of $T_0$ and $\rho$ is density. \citet{2005ApJ...619..538R} conclude that a value of $T_0 = 10^6$\,K can reproduce the Spite Plateau in Population II stars. We have therefore adopted this value. Note that there is (as yet) no physical cause for such mixing and the prescription remains an {\it ad hoc} one. 

\begin{figure}
\includegraphics[width=\columnwidth]{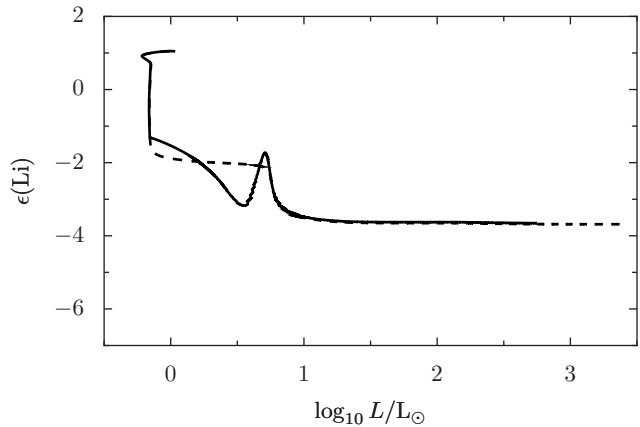}
\caption{The evolution of the lithium abundance as a function of luminosity when 0.1\ms\ of material is accreted. The solid line shows the case where thermohaline mixing and gravitational settling are included. The dashed line has the additional turbulent mixing of \citet{2005ApJ...619..538R} included.}
\label{fig:extra}
\end{figure}

We have recomputed the 0.1\ms\ case with this additional mixing mechanism included (in addition to thermohaline mixing and gravitational settling). The results are shown in Figure~\ref{fig:extra}. The addition of extra mixing reduces the height of the $\mu$-barrier and hence allows somewhat more efficient thermohaline mixing. The degree of Li depletion across the main sequence is significantly reduced because the additional mixing inhibits gravitational settling by stirring up the surface regions. The extra mixing does not substantially change the Li abundance however, as extensive mixing can still take place and it is this that causes the destruction of Li.

\subsection{Fluorine}

\begin{figure}
\includegraphics[width=\columnwidth]{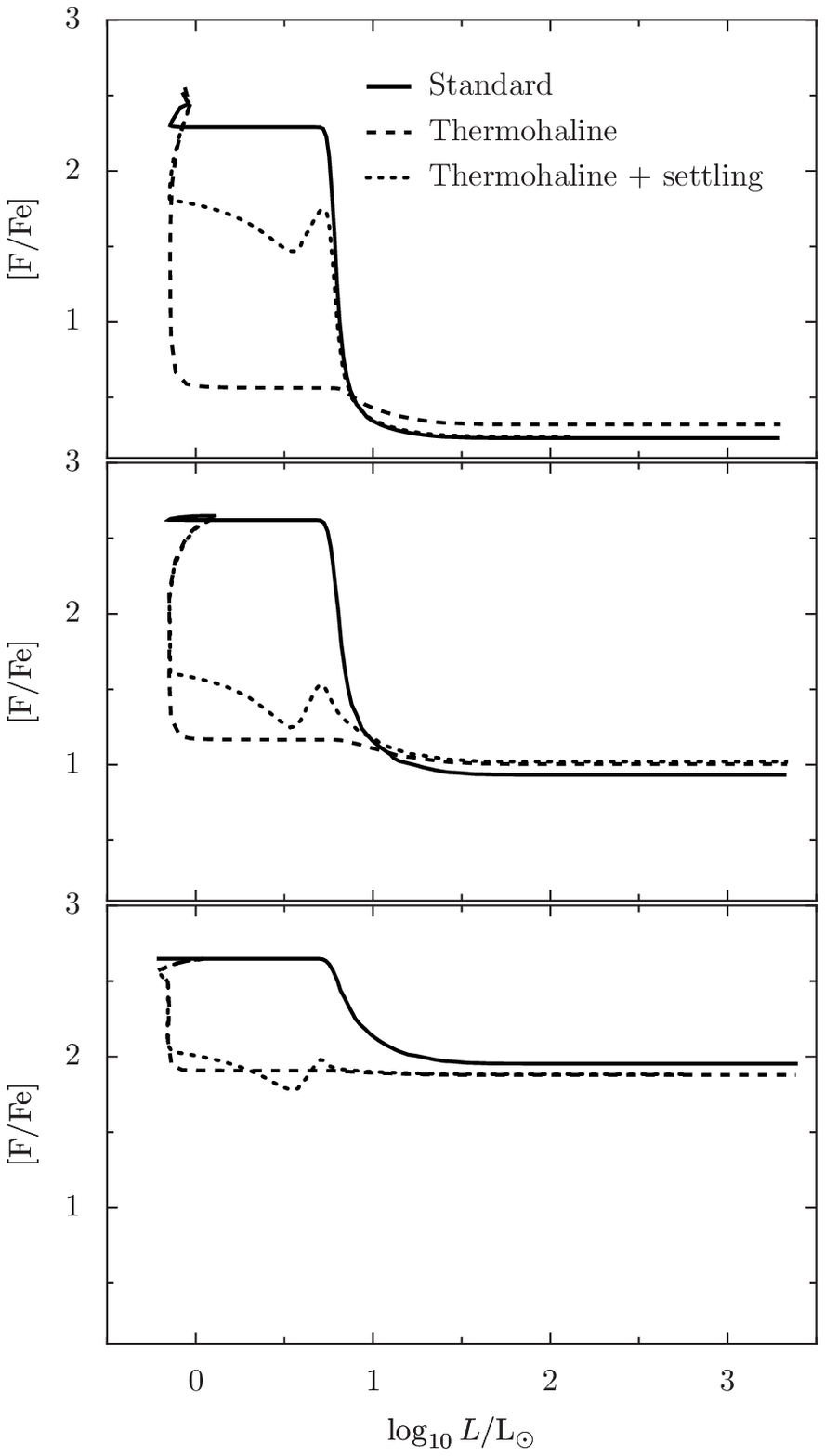}
\caption{The evolution of [F/Fe] as a function of luminosity for each of the model sequences. The standard models are displayed with solid lines and those with only thermohaline mixing are displayed with dashed lines. The models with both thermohaline mixing and gravitational settling are shown by dotted lines. From top to bottom, the panels represent the case of accreting 0.002, 0.01 and 0.1\ms\ of material respectively. In each case, material has been accreted from a 1.5\ms\ companion and the final mass of the star is 0.8\ms.}
\label{fig:FStack}
\end{figure}

The evolution of the fluorine abundance for a selection of the models is showing in Figure~\ref{fig:FStack}. Without any additional mixing mechanism, the secondary retains the surface abundance of the AGB ejecta which is [F/Fe]~$\approx2.3$. It then suffers a dilution at first dredge-up which, depending on the mass of material accreted, can cause [F/Fe] to drop by between 0.5-2 dex. Models including thermohaline mixing have an immediate dilution of fluorine of around this level and may also suffer a further dilution at first dredge-up, depending on whether the accreted material is mixed to a depth greater than the depth of the convective envelope during first dredge-up. In the case that both thermohaline mixing and gravitational settling are included, the fluorine abundance suffers less of a reduction on the main sequence, but only if a small quantity of material is accreted. A surface abundance of [F/Fe] of around 2 or less is predicted. This value declines slightly over the main sequence due to the effects of gravitational settling.

Fluorine has only been measured in one CEMP star to date: HE 1305+0132. This object has an extremely supersolar F abundance of [F/Fe]~$=~+2.90$ \citep{2007ApJ...667L..81S}. The metallicity is [Fe/H]~=~-2.5 and the star does show barium lines in its spectrum \citep{2005MNRAS.359..531G} and so we should expect the model presented here to be appropriate. HE~1305+0132 is a red giant with a surface gravity of $\log_{10} g=0.8$ and hence we should compare this to the models once they have passed through first dredge-up. Unfortunately, we are unable to distinguish between the three model sequences once the star has passed through first dredge-up as they all give similar [F/Fe] values. However, it is difficult to reconcile the models with the observations. There is insufficient fluorine in the AGB ejecta to start with as the models do not reach high enough [F/Fe] on the main sequence, let alone after dilution of material has taken place either via thermohaline mixing or during first dredge-up. The case is worse if less material is accreted. If 0.1\ms\ of material is accreted the models are about 1 dex below the observed [F/Fe] of HE~1305+0132, while if only 0.001\ms\ are accreted the discrepancy is over 2 dex. However, the [C/Fe] and [N/Fe] values are quite well reproduced by a model that has accreted 0.1\ms\ of material from a 1.5\ms\ companion and mixed this material during the main sequence (this is necessary to elevate the nitrogen abundance, which happens via CN-cycling as describe by \citet{2007A&A...464L..57S}.). On the giant branch, this model has [C/Fe]$\ = 2.0$ and [N/Fe]$\ = 1.5$, which compare very favourably with the measured values of 2.2 and 1.6 respectively \citep{2007ApJ...667L..81S}. 

The only way out of this dilemma is to suggest that the AGB models are underabundant in fluorine. There could be two causes for this: the model has a companion of a too low a mass to produce enough fluorine, or there is an intrinsic problem with the production of fluorine in the AGB models. On the former point, \citet{2008A&A...484L..27L} have shown that a 2.25\ms\ star is likely to produce the maximum fluorine yield. On the latter point, the nucleosynthesis of fluorine takes place via an involved chain of reactions \citep[see][for details]{2004ApJ...615..934L}. There are many uncertainties associated with the production and destruction of \el{19}{F} \citep[see the discussion in][]{2008ApJ...676.1254K}. On top of this, there are model uncertainties such as the efficiency of third dredge-up \citep[e.g.][among many others]{1996ApJ...473..383F,2000A&A...360..952H,2005MNRAS.356L...1S} and the mass loss rates used \citep[e.g.][]{2007MNRAS.375.1280S}. As an indication of such effects, we note that the models of \citet{2007PASA...24..103K} are more abundant in \el{19}{F} than the SG08 models used here, by roughly an order of magnitude at 2\ms. This may allow the reconciliation of the observations with theoretical models, assuming a substantial quantity of material was accreted.

\subsection{Sodium}

\begin{figure}
\includegraphics[width=\columnwidth]{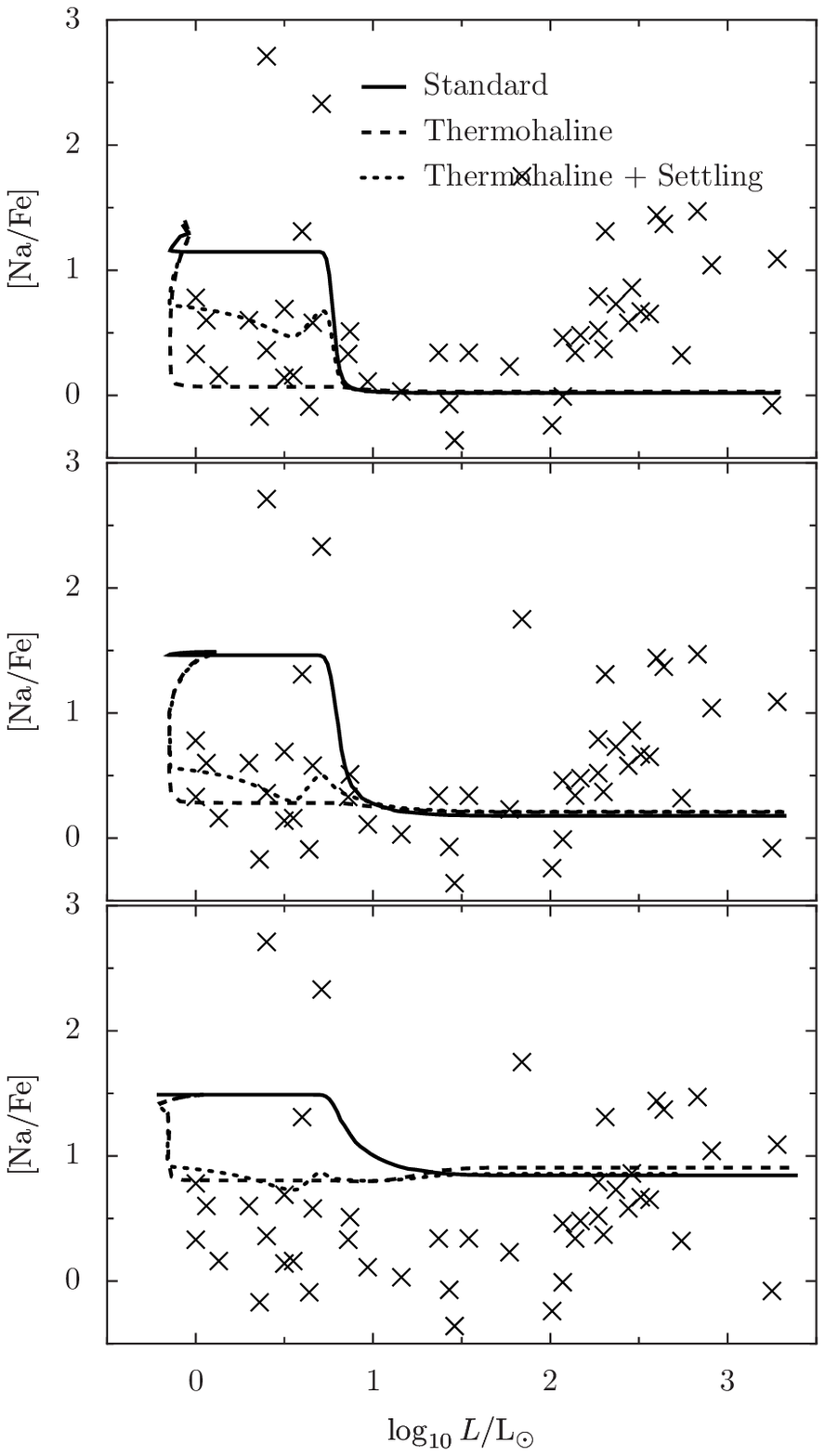}
\caption{The evolution of [Na/Fe] as a function of luminosity for each of the model sequences. The standard models are displayed with solid lines and those with only thermohaline mixing are displayed with dashed lines. The models with both thermohaline mixing and gravitational settling are shown by dotted lines. From top to bottom, the panels represent the case of accreting 0.002, 0.01 and 0.1\ms\ of material respectively. In each case, material has been accreted from a 1.5\ms\ companion and the final mass of the star is 0.8\ms. Crosses represent the Ba-rich stars of \citet{2007ApJ...655..492A} and \citet{2008ApJ...678.1351A}. Typical random errors on the abundance determinations are between 0.05 and 0.15 dex \citep{2008ApJ...678.1351A}.} 
\label{fig:NaStack}
\end{figure}

The evolution of the sodium abundance as a function of luminosity for models accreting material from a 1.5\ms\ companion is shown in Figure~\ref{fig:NaStack}. The figure also displays the barium-enriched CEMP stars from \citet{2007ApJ...655..492A} and \citet{2008ApJ...678.1351A} that have measured sodium abundances. Most of the observed stars have sodium abundances in the range [Na/Fe]~$=~0-1$, though there are some extreme outliers which are highly enhanced in sodium, with [Na/Fe] around 2 or greater.

The sodium abundance is best matched by models that include mixing (either by thermohaline mixing alone, or by the combination of thermohaline mixing and gravitational settling). The sodium abundance is relatively flat and shows no step change at or around first dredge-up. If accreted material remained unmixed during the main sequence, we would expect to see a drop of between 0.5 and 1 dex in [Na/Fe] at $\log\ L/\mathrm{L_\odot}\approx 0.7$. The data do not support a drop of this magnitude, though we caution that the data are a combination of a spread in metallicity and presumably also in accreted mass and companion mass. However, the ejected material of the 1.5\ms\ model is very sodium rich, with [Na/Fe]~$\ \approx~1.5$ and any unmixed material should stand out. The models with thermohaline mixing bracket the bulk of the observations quite well, with the model where just 0.002\ms\ is accreted matching the lower observations while the model where 0.1\ms\ is accreted lies towards the upper observations.

An alternative explanation is that the observed systems all have accreted material that is less sodium rich than our model. This would suggested that they all (excluding the extremely Na-rich outliers) had companions of lower mass than 1.5\ms. This is possible but it would require the majority of these binary systems to have mass ratios close to unity. If this is the case, it would be difficult to draw any conclusion about mixing from the sodium abundances.

One may question whether the initial abundances are appropriate. If we look at the metal-poor but carbon-normal stars in the SAGA database \citep{2008arXiv0806.3697S}, we find a base level of sodium to be around [Na/Fe]~=~0.5, though there is evidence that this increases with decreasing metallicity and there is some scatter. This is comparable to the [Na/Fe] values observed in the CEMP stars. This reinforces the need for mixing as the AGB material is considerably more sodium enhanced than this, by about one dex. 

What about those objects with enhanced sodium? The two most sodium-rich objects in the sample are CS~29528-028 which has [Na/Fe]~=~2.33 \citep{2007ApJ...655..492A} and SDSS~1707+58 which has [Na/Fe]~=~2.71 \citep{2008ApJ...678.1351A}. It is possible that these two objects had companions that were more massive than the 1.5\ms\ companions considered so far. In order to produce sodium during the AGB phase of the primary, we need first to make \el{22}{Ne}. This is produced in the intershell via the reactions:
\be
\el{14}{N}\ag\el{18}{O}\ag\el{22}{Ne}.
\ee
The addition of a proton to this \el{22}{Ne} produces \el{23}{Na} and this occurs in the hydrogen buring shell. The more massive AGB stars undergo more thermal pulses and thus can dredge up more \el{22}{Ne}. They also tend to have deeper convective envelopes, which can lead to more \el{23}{Na} production via hot bottom burning.

We have also run the model sequences accreting material from companions of 2 and 3.5\ms, using the ejecta abundances as computed by SG08 (some of which are displayed in Table~\ref{tab:abundances}). The ejecta from the latter model has [Na/Fe]~=~2.50, which is greater than the observed value for CS~29528-028, but lower than that for SDSS~1707+58 by about 0.2 dex. Thus we can reproduce the observed sodium abundances, but only if the accreted material is not mixed into the secondary. The carbon and nitrogen abudances of the 3.5\ms\ model give [C/Fe]~=~2.18 and [N/Fe]~=~3.10. These are in reasonable agreement with the measured [C/Fe] and [N/Fe] values in CS~29528-028, which are 2.77 and 3.58 respectively. The magnesium abundance is also in reasonable agreement as well (see below). It seems plausible that this object was once the companion to an AGB star of around 3.5\ms. We are forced to assume that the accreted material remained unmixed with the star, unless an order of magnitude more of each element were present in the ejecta. This seems unlikely as discussed in Section~\ref{sec:discussion}.

The data points appear to show that the sodium abundance begins to rise once the luminosity reaches $\log L/\mathrm{L_\odot}\approx2$. Prior to this, [Na/Fe] is around 0.4 while after this point there is a marked increase. This trend should be treated with caution as there are few data points, but it may point to the existence of extra mixing on the red giant branch. This would require a non-canonical mixing mechanism that could reach a stellar layer deep enough for the NeNa cycle to be active. Proton captures on to \el{22}{Ne} begin at a temperature of around $2\times10^7$\,K \citep{1999A&A...347..572A}.  The ejecta of the 1.5\ms\ model is rich in \el{22}{Ne} and so there is enough raw material to produce the necessary sodium.

Let us assume that the extra mixing that occurs on the giant branch is via the \el{3}{He} mechanism of \citet{2006Sci...314.1580E} and that we can reproduce the effects using thermohaline mixing following the work of \citet{2007A&A...467L..15C}. In order to do this, we need to increase the value of the thermohaline mixing coefficient by about two orders of magnitude above the \citet{1980A&A....91..175K} value. We have re-run the post-accretion sequence of the case in which 0.1\ms\ of material is accreted from a 1.5\ms\ companion. We find that while an increase in [Na/Fe] does occur, it is of an insignificant amount (about 0.04 dex). In addition, there is a depletion of [C/Fe] at the level of about 0.1 dex and an increase in [N/Fe] of about 0.2 dex. The data do not support the occurrence of extra mixing in CEMP stars on the upper part of the giant branch \citep[see][for further discussion]{2008ApJ...679.1541D}. Given the spread in the nitrogen data is of the order of 1-1.5 dex, it would be easy to hide such a change in among the noise. 

Can we enhance this process? Enhancing the rate at which thermohaline mixing occurs in the giant branch is possible, but likely to lead to the undesirable consequences that carbon is substantially depleted and nitrogen substantially enhanced. There is no observational evidence that this happens. Also, the work of \citet{2007A&A...467L..15C} seems to indicate that the efficiency of mixing we are using is correct. So how do we produce more \el{23}{Na}? For a given temperature and density, we can increase the rate of production by increasing the abundance of \el{22}{Ne} (this could be caused by having more efficient TDUP and/or more thermal pulses with TDUP). This element is one of the most abundant in the ejecta from an AGB star and it will not become severely depleted by p-burning reactions. A star which contains more \el{22}{Ne} in its envelope after first dredge-up will produce a greater amount of sodium during extra mixing towards the tip of the giant branch.

\begin{figure}
\includegraphics[width=\columnwidth]{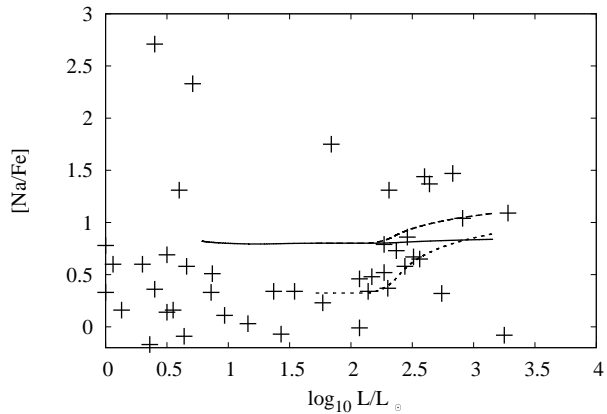}
\caption{The evolution of the surface [Na/Fe] abundance as a function of luminosity for the stellar models. The solid line displays the model with enhanced thermohaline mixing. The dashed line shows the model in which the the \el{22}{Ne} abundance in the envelope has been increase by a factor of 10. The dotted line shows the model with reduced \el{23}{Na} abundance and is a better representation of the data. Crosses represent the Ba-rich stars of \citet{2007ApJ...655..492A} and \citet{2008ApJ...678.1351A}. Typical random errors on the abundance determinations are between 0.05 and 0.15 dex \citep{2008ApJ...678.1351A}.}
\label{fig:NaEnhanced}
\end{figure}

To test this we have taken one of our models after it has passed through first dredge-up and artificially enhanced the \el{22}{Ne} abundance in the envelope by a factor of 10. Boosting the abundance of \el{22}{Ne} by such a large amount is somewhat extreme but within the variation of model predictions from different codes, which can vary by an order of magnitude (see e.g. figure 2 in SG08). The model is then evolved to the tip of the giant branch. We have taken the model from the sequence where 0.1\ms\ of material from a 1.5\ms\ companion has been accreted. Consequently the material is somewhat \el{23}{Na}-rich. The results are shown in Figure~\ref{fig:NaEnhanced}. We note that we can get an upturn in [Na/Fe] at the correct luminosity and a substantial increase in the abundance of \el{22}{Ne} in the accreted material does allow [Na/Fe] to a level  comparable to that observed (the dashed line in Figure~\ref{fig:NaEnhanced}). If we reduce the Na abundance to a value that better represents the [Na/Fe] abundance after first dredge-up (i.e. we assume we have not accreted as much material\footnote{We must still satisfy the requirement [C/Fe]$\ >+1$ and this will indeed be met provided that the accreted mass is no less than about 0.01\ms. Note that this is roughly the minimum accreted mass required for a {\it giant} to appear carbon-rich as in such a star the accreted material has at least been diluted during first dredge-up (dilution could have happened earlier, via thermohaline mixing). A main-sequence object whose accreted material has not been mixed may appear C-rich even if it has accreted far less material, as can be seen in figure 4 of SG08.}, or the material is not as Na-rich as the AGB model predicts), we get much better agreement with the data. We stress that this is a simple test based on arbitrarily changing the Ne and Na abundances. Whether a stellar model (or indeed a real star) can produce the necessary abundances is another matter entirely.

\subsection{Magnesium}

\begin{figure}
\includegraphics[width=\columnwidth]{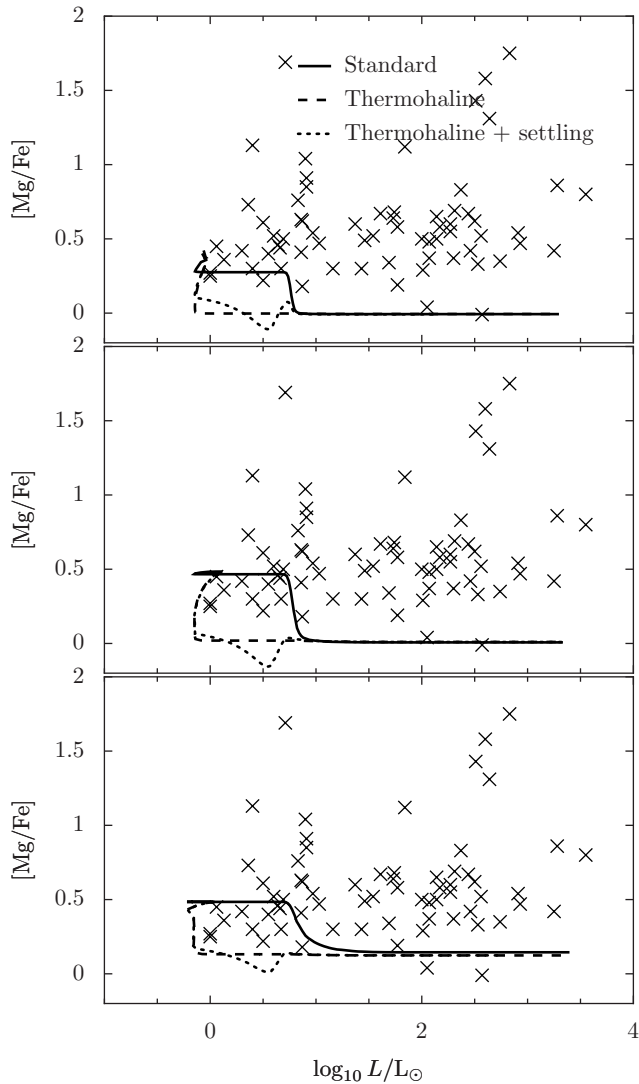}
\caption{The evolution of [Mg/Fe] as a function of luminosity for each of the model sequences. The abundance is the sum of the three stable magnesium isotopes \el{24}{Mg}, \el{25}{Mg} and \el{26}{Mg}. The standard models are displayed with solid lines and those with only thermohaline mixing are displayed with dashed lines. The models with both thermohaline mixing and gravitational settling are shown by dotted lines. From top to bottom, the panels represent the case of accreting 0.002, 0.01 and 0.1\ms\ of material respectively. In each case, material has been accreted from a 1.5\ms\ companion and the final mass of the star is 0.8\ms. Crosses represent the Ba-rich stars of \citet{2007ApJ...655..492A} and \citet{2008ApJ...678.1351A}. Typical random errors on the abundance determinations are between 0.05 and 0.15 dex \citep{2008ApJ...678.1351A}.}
\label{fig:MgStack}
\end{figure}

The evolution of the surface magnesium abundance as a function of luminosity for models accreting 0.002, 0.01 and 0.1\ms\ of material from a 1.5\ms\ companion are shown in Figure~\ref{fig:MgStack}. The abundance plotted is the sum of the three stable magnesium isotopes \el{24}{Mg}, \el{25}{Mg} and \el{26}{Mg}. Of these, the AGB material is mostly enhanced in \el{25}{Mg} (with slightly less \el{26}{Mg}) as a result of the activation of the \el{22}{Ne}\an\el{25}{Mg} reaction in the intershell and the subsequent dredge-up of this material. Of the displayed models, only the models without thermohaline mixing pass through the observed data points and this only occurs while the stars are on the main sequence. Once first dredge-up has occurred and the accreted material has become mixed, the models all skirt the bottom edge of the observations.

This discrepancy is readily explained by our choice of solar-scaled initial abundances. In low metallicity environments, stars appear to be enhanced in $\alpha$-elements and we could reasonably expect [Mg/Fe] to be around 0.3 at this metallicity \citep[e.g.][]{2004A&A...416.1117C,2004ApJ...612.1107C}. This is presumably only due to an enhancement of the $\alpha$-element \el{24}{Mg}. As this isotope undergoes only minor processing during the AGB (at least for low-mass stars like the 1.5\ms\ case considered here), an increase in the initial abundance of \el{24}{Mg} would shift the tracks to higher [Mg/Fe] and also reduce the magnitude of the dilution (irrespective of whether the dilution is caused by thermohaline mixing on the main sequence or convective mixing during first dredge-up) which would provide better agreement with the data.

\begin{figure}
\includegraphics[width=\columnwidth]{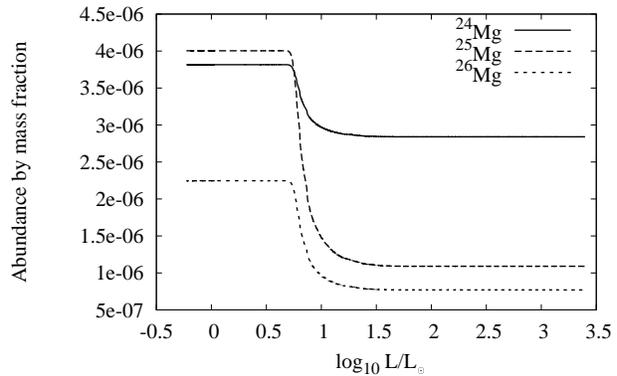}
\caption{The evolution of the magnesium isotopes as a function of luminosity for the model in which 0.1\ms\ of material is accreted from a 1.5\ms\ companion and only canonical (i.e. convective) mixing processes are included. The tracks begin from the end of the accretion phase. The abundances (by mass fraction) of \el{24}{Mg}, \el{25}{Mg} and \el{26}{Mg} in the pristine stellar material are $2.59\times10^{-6}$, $3.40\times10^{-7}$ and $3.90\times10^{-7}$ respectively. Note that \el{25}{Mg} is the dominant isotope of Mg in the accreted material (by a very small margin) prior to first dredge-up, which occurs between $\log_{10} L/\mathrm{L_\odot}$ of around 0.7 to 1.5.}
\label{fig:MgAbund}
\end{figure}

Figure~\ref{fig:MgAbund} demonstrates this point. In the accreted material, \el{25}{Mg} is the dominant magnesium isotope, followed by \el{24}{Mg} (the ratios of the magnesium isotopes in the AGB ejecta are \el{25}{Mg}/\el{24}{Mg}$=1.05$ and \el{26}{Mg}/\el{24}{Mg}$=0.59$.). Because \el{24}{Mg} is little affected on the AGB, the abundance in the accreted material is closer to that of the pristine matter. When the accreted material is mixed into the secondary, the surface abundance of \el{24}{Mg} suffers less of a depletion than that of the \el{25}{Mg}. The \el{24}{Mg} abundance changes by about a quarter, while the \el{25}{Mg} abundance is reduced by a factor of 4. We would expect that if we boosted the abundance of \el{24}{Mg}, it would dominate the total magnesium abundance and lead to a much reduced change in [Mg/Fe] when the material becomes mixed.

\begin{figure}
\includegraphics[width=\columnwidth]{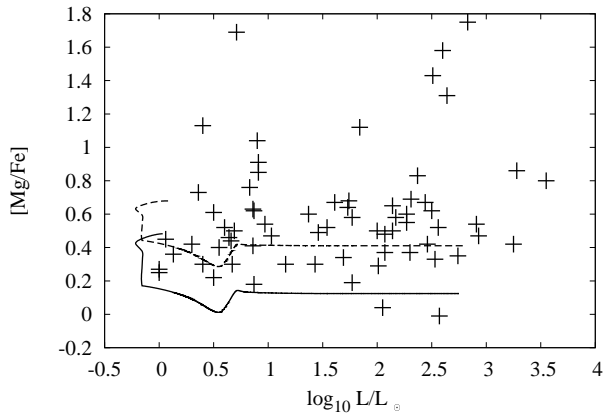}
\caption{The evolution of the surface [Mg/Fe] ratio as a function of luminosity when 0.1\ms\ of material from a 1.5\ms\ companion is accreted. Both thermohaline mixing and gravitational settling are included. The solid line displays the standard case while the dashed line is for the model with an enhanced \el{24}{Mg} abundance in both the secondary and the accreted material.}
\label{fig:MgEnhanced}
\end{figure}

In order to test this, we have re-run the case of accretion of 0.1\ms\ of material from a 1.5\ms\ companion, with thermohaline mixing and gravitational settling included. In this model, we have arbitrarily increased the \el{24}{Mg} abundance in the AGB ejecta and in the secondary by a factor of 2.5. This gives the secondary a magnesium-to-iron abundance of [Mg/Fe]~=~+0.4, which is comparable to that of carbon-normal EMP stars. In this model, the accreted material now has [Mg/Fe]~=~0.68 and upon mixing, an equilibrium value of [Mg/Fe]~=~0.4 is restored. The model now passes through the middle of the data, providing much better agreement, as shown in Figure~\ref{fig:MgEnhanced}.

There is little evidence for any variation in the magnesium abundance as a function of luminosity, though there is considerable spread in the data. This does not sit well with stellar models that do not include thermohaline mixing of the accreted material, as these would be expected to show a step-change in the Mg abundance at first dredge-up owing to the dilution of accreted material at this point. A step-change of 0.5 dex, as seen in the original solar-scaled standard models (i.e. those with convective mixing only), ought to be detectable among the spread in the data. However, in an $\alpha$-enhanced model the drop in the Mg abundance that occurs at first dredge-up would be reduced, as there is more \el{24}{Mg} in the pristine material. Consequently, the drop in [Mg/Fe] at first dredge-up would become comparable to the scatter in the observations and therefore may not be evident.

There are several objects that are extremely magnesium rich. The 4 most Mg-rich are: CS~29498-043, CS~29528-028, HE~1447+0102 and HE~0039-2635. We can reconcile the observations of CS~29528-028 with an object that has accreted material from a 3.5\ms\ companion, but we have to assume the material remains unmixed on the main sequence. This object has [Mg/Fe]~=~1.69, compared to [Mg/Fe]~=~1.52 in the AGB ejecta.  The other, more luminous, stars -- which must have undergone first dredge-up -- are difficult to explain. This is discussed in more detail in Section~\ref{sec:discussion}. Here we simply note that a star that has passed through first dredge-up must necessarily have undergone some sort of extensive dilution, even if the accreted material had remained unmixed on the stellar surface during the main sequence. This would require the accreted material of have an abundance of magnesium that is considerably higher than the 3.5\ms\ model and presumably a large amount of material would have to be accreted to minimise the effects of dilution. This in not unfeasible: \citet{2007PASA...24..103K} have a total magnesium abundance (which is dominated by \el{26}{Mg}) of around $10^{-3}$ by mass fraction in their 3\ms, Z=0.0001 stellar model. Note that the required abundance in the accreted material increases as the mass of the accreted material decreases. 

It must be pointed out that even though an AGB model undergoes hot bottom burning, it can still be {\it both} a CEMP star and a nitrogen-enhanced metal-poor star (NEMP). NEMPs are defined by \citet{2007ApJ...658.1203J} as having [N/Fe]$\ > +0.5$ and [C/N]$\ <-0.5$. The star CS~29528-028 has [C/Fe]$\ = 2.77$ and [N/Fe]$\ = 3.58$ making it both a CEMP and a NEMP star. The ejecta of the 3.5\ms\ model of SG08 has [C/Fe]$ = 2.18$ and [N/Fe]$\ = 3.19$ which makes a reasonable match to CS~29528-028's abundances.

As a caveat to the above, we note that we have assumed the iron abundance is [Fe/H]~=~-2.3, consistent with the stellar models used. In fact, CS~29498-043 is considerably more metal-poor than this, with [Fe/H]~=~-3.54 \citep{2007ApJ...655..492A}. This would reduce the Mg abundance required in the ejecta considerably. However, the other objects are of a comparable metallicity to the models presented here so this does not affect our conclusion on the magnesium enhancements of the accreted material.

\section{Discussion}\label{sec:discussion}

\begin{figure}
\includegraphics[width=\columnwidth]{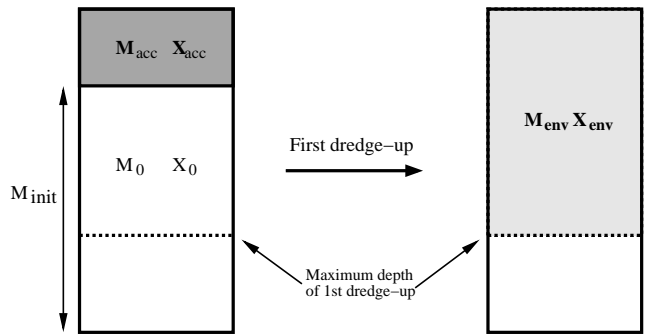}
\caption{Schematic representation of mixing after accretion. A star of mass $M_\mathrm{init}$ and initial composition $X_0$ accretes $M_\mathrm{acc}$ of material of composition $X_\mathrm{acc}$ from its companion. After first dredge-up, the envelope (of mass $M_\mathrm{env} = M_0 + M_\mathrm{acc}$) has become mixed to a homogenous compostion of $X_\mathrm{env}$.}
\label{fig:DUPmix}
\end{figure}

We can do a simple calculation to determine the level of enrichment required to match the observed constraints. In the simplest case, we assume that there is no burning (i.e. the abundances are unaffected by extra mixing on the giant branch) of material and that material is mixed during first dredge-up only. The abundance in the envelope after mixing is determined by the depth of first dredge-up. At the deepest point of envelope penetration, the mass of a given isotope contained within the envelope is just $X_\mathrm{env}M_\mathrm{env}$ where $X_\mathrm{env}$ is the mass fraction of the isotope in the envelope and $M_\mathrm{env}$ is the mass in the envelope. By conservation of mass, this must be equal to the sum of the mass of that isotope in the accreted matter $X_\mathrm{acc}M_\mathrm{acc}$ plus the mass of that isotope contained in the original star, down to the maximum point of envelope penetration, $X_0M_0$ (see Figure~\ref{fig:DUPmix} for a schematic representation). This can be re-arranged to give:
\be
X_\mathrm{acc} = {X_\mathrm{env}M_\mathrm{env} - X_0M_0\over M_\mathrm{acc}},
\ee
which we may use to work out the required abundance of a given isotope in the AGB ejecta which will give the observed abundance on the giant branch, given an assumed mass of accreted material. Stellar models (e.g. SG08, \citealt{2008A&A...484L..27L}) suggest that the convective envelope penetrates to a depth of around $M_\mathrm{env}=0.45$\ms\ in a star of 0.8\ms\ at $Z=10^{-4}$.

In the case of the fluorine-rich star HE 1305+0132 which has [F/Fe]~=~+2.9 and [Fe/H]~=~-2.5, the observed abundance of fluorine implies a mass fraction of $X_\mathrm{F}=1.65\times10^{-6}$. Using the above formula with a solar-scaled fluorine abundance yields $X_\mathrm{acc} = 7.41\times10^{-4}$ if we assume only 0.001\ms\ of material is accreted, while this value is one hundred times lower if we accrete 0.1\ms\ of material. The latter case is almost reconcilable with the SG08 models, which give $X_\mathrm{F}\approx10^{-6}$ for stellar models of 1.5 and 2\ms. The models of \citet{2007PASA...24..103K} and \citet{2008A&A...484L..27L} give about an order of magnitude more fluorine giving better agreement with the observations. 

The occurrence of thermohaline mixing does not substantially affect the fluorine abundance required in the AGB ejecta. On the assumption that material mixes to the maximum depth predicted by SG08 (i.e. around 0.7\ms\ of the star becomes mixed), the required mass fraction in the accreted material changes by less than a factor of two. The high fluorine abundance associated with only a small amount of accreted material is difficult to reconcile with stellar models and it therefore seems likely that this object received a considerable quantity of AGB material. Similar conclusions were also reached by \citet{2008A&A...484L..27L}, who suggest that this object probably received between 3-11 per cent (0.05-0.12\ms) of the material ejected by a 2\ms\ companion. These authors also point out that such extremely F-rich CEMP stars ought to be rare, though F-enhancement likely occurs alongside C- and $s$-processes enhancements and should be sought out as a means of confirming the AGB mass transfer scenario.

Repeating this analysis for the highest observed sodium and magnesium enrichments, we find required abundances vary from $X_\mathrm{Na} = 3\times10^{-4} - 10^{-2}$ and $X_\mathrm{Mg} = 8\times10^{-4} - 10^{-2}$. Again, the high abundances are irreconcilable with stellar models. It seems highly unlikely that stellar models would produce sodium or magnesium at mass fractions of around 1 per cent! Current stellar models predict the upper limit on the abundances of these two elements to be in the range $10^{-4}-10^{-3}$ \citep[e.g.][and SG08]{2004ApJS..155..651H, 2007PASA...24..103K}, depending on the input physics and the mass of the stellar model. This upper limit is roughly consistent with our simple calculations for the required sodium and magnesium abundances in the accreted material. It is tempting to suggest that many CEMP stars do not accrete a great deal of material from their companions, on the grounds that we do not see many highly enriched objects (regardless of whether that enrichment is in terms of F, Na or Mg).

The models computed in this work have not included the effects of radiative levitation. This would be expected to temporarily raise the surface abundance of some of the heavy elements, particularly Na and Mg \citep[see figure 4 in][]{2002ApJ...580.1100R}. It would also raise the Fe abundance, so the exact effect on the ratio [X/Fe] is not straightforward to predict. Radiative levitation will not change the qualitative picture presented herein because it is only effective over long (i.e. Gyr) timescales just like gravitational settling. It does not substantially affect the settling of helium and hence the $\mu$-barrier that may develop. As it is this that determines the extent to which thermohaline mixing can occur and because thermohaline mixing operates over a short timescale relative to the settling processes, the qualitative picture presented herein should not change. In addition, if an extra turbulent mixing process is needed to explain the differences in the iron and magnesium abundances that exist between turn-off and evolved stars in globular clusters \citep{2008A&A...490..777L}, the omission of radiative levitation will have less of an effect as extra mixing reduces the impact of settling and levitation. The predictions for the giant branch will be unchanged because the material in the envelope is thoroughly mixed by convection: it is only the main-sequence predictions that will be affected.

A wealth of information is also available for several heavy species not included in our network. Several authors have compared the $s$-process element abundance trends to predictions to stellar models and come up with factors reflecting the degree of dilution that must have occurred in the star \citep[e.g.][]{2008ApJ...677..556T, 2008ApJ...679.1549R}. In particular, comparing these heavy metal abundance patterns in turn-off stars to model predictions may help to elucidate the nature and extent of any mixing on the main-sequence. We intend to pursue this avenue in future work.

The work discussed herein only applies to those CEMP stars that have formed via the accretion of material from an AGB companion. This likely applies to the CEMP-$s$ stars, which account for the majority of the CEMP population \citep[around 70 per cent, according to][]{2003IAUJD..15E..19A}. The remaining CEMP stars are likely to have formed via a different mechanism. Other proposed formation scenarios invoke the ejecta of rapidly rotating massive stars \citep{2006A&A...447..623M} or faint type II supernovae \citep{2005ApJ...619..427U}. It is also possible for extremely metal-poor stars to enrich themselves through a dual core flash, where protons are ingested into the convective region that is driven during the core He-flash at the tip of the giant branch \citep[e.g.][]{1990ApJ...349..580F}. This only happens for stars with metallicities below [Fe/H]$\ \leq -5$ \citep{2008A&A...490..769C}, which is considerably more metal-poor than the objects discussed in this work. 

\section{Conclusion}

The measurement of a high Li abundance in an unevolved CEMP star precludes the possibility of {\it extensive} mixing of accreted material. This could either be because there is no thermohaline mixing or because only a small quantity of material is accreted (likely a few thousandths of a solar mass). The inclusion of gravitational settling does not substantially increase the very stringent limit imposed on the amount of mass that can be accreted. It is possible that the Li abundance in accreted AGB material is substantially higher than computed at present owing to the occurrence of some additional mixing mechanism during the thermally pulsing asymptotic giant branch. However, the Li enhancements required to make the thermohaline mixing models fit the observations  would have to be at least two orders of magnitude above the current yield. This seems an unlikely level of enhancement.

Comparing measured sodium abundances to the models suggests that the majority, {\it but not all}, stars undergo mixing during the main sequence. The abundance in the AGB ejecta is significantly higher than that observed in most CEMP stars. Unless sodium is produced to a much lower degree in AGB stars than the models predict (or the stars we are observing all had companions of much lower mass than we have assumed), it is difficult to reconcile models without main-sequence mixing with the observed abundances. However, it is difficult to explain the extremely sodium-rich objects with models that {\it do} include thermohaline mixing as this would require the accreted material to have a sodium abundance considerably higher than current model predictions. It therefore seems prudent to argue that some stars mix their accreted material while others do not or that there is a spread in the efficiency of the mixing of accreted material. The cause for this is unclear and merits investigating. One possible suggestion is that rotation may play a role. \citet{2008ApJ...684..626D} have pointed out that rotationally-driven horizontal turbulence may suppress thermohaline mixing. If the secondary stars are rotating at different rates (or the angular momentum content of the accreted material varies from star-to-star) then some stars may experience thermohaline mixing whilst others do not. 

The flatness of the observed magnesium abundances also does not support models without thermohaline mixing on the main sequence. However, this may be because the models used are not $\alpha$-enhanced and consequently are likely to be less rich in \el{24}{Mg} than the observed stars. AGB and secondary models with $\alpha$-enhanced compositions and abundances should certainly be investigated. As the magnesium from AGB stars is predominantly \el{25,26}{Mg}, additional \el{24}{Mg} in the stellar material could easily mask the dilution of the AGB material and hence we would not be able to see any changes due to mixing events. However, if the CEMP stars have received a substantial quantity of magnesium from an AGB companion, it is likely to be in the form of the neutron-rich isotopes \el{25}{Mg} and \el{26}{Mg}. Hence the signature of accretion from such a companion may be detectable in the ratios of \el{25,26}{Mg} to \el{24}{Mg}. Such measurements have been made in metal-poor but carbon-normal stars but the results are inconclusive \citep{2003ApJ...599.1357Y,2007ApJ...659L..25M}. 

The cases of extreme enhancement of F, Na and Mg on the giant branch argue against the accretion of only small quantities of material (i.e. thousandths of a solar mass) in these cases. The required abundances in the AGB ejecta would be beyond the current range of model predictions. The accretion of much greater amounts of material would be expected to show more extensive mixing on the main sequence if thermohaline mixing is efficient.

No clear picture arises from the elements studied here. The detection of lithium in turn-off objects seems to rule out extensive mixing for these cases. Some highly sodium and magnesium enriched turn-off objects can only be explained if accreted material remains unmixed during the main sequence. However, measurements of [Na/Fe] seems to suggest that most CEMP stars do efficiently mix their accreted material. A more comprehensive study involving a greater number of elements may help to elucidate what processes go on in these objects.

\section{Acknowledgements}
The anonymous referee is thanked for their useful comments which have helped to improve the manuscript. RJS thanks Ross Church for his careful reading of the manuscript prior to its submission. The author is funded by the Australian Research Council's Discovery Projects scheme under grant DP0879472. 

\bibliography{../../../masterbibliography}

\begin{thebibliography}{}

\bibitem[\protect\citeauthoryear{{Abia} \& {Isern}}{{Abia} \&
  {Isern}}{1997}]{1997MNRAS.289L..11A}
{Abia} C.,  {Isern} J.,  1997, MNRAS, 289, L11

\bibitem[\protect\citeauthoryear{{Aoki}, {Beers}, {Christlieb}, {Norris},
  {Ryan} \& {Tsangarides}}{{Aoki} et~al.}{2007}]{2007ApJ...655..492A}
{Aoki} W.,  {Beers} T.~C.,  {Christlieb} N.,  {Norris} J.~E.,  {Ryan} S.~G.,
  {Tsangarides} S.,  2007, ApJ., 655, 492

\bibitem[\protect\citeauthoryear{{Aoki}, {Beers}, {Sivarani}, {Marsteller},
  {Lee}, {Honda}, {Norris}, {Ryan} \& {Carollo}}{{Aoki}
  et~al.}{2008}]{2008ApJ...678.1351A}
{Aoki} W.,  {Beers} T.~C.,  {Sivarani} T.,  {Marsteller} B.,  {Lee} Y.~S.,
  {Honda} S.,  {Norris} J.~E.,  {Ryan} S.~G.,    {Carollo} D.,  2008, ApJ, 678,
  1351

\bibitem[\protect\citeauthoryear{{Aoki}, {Ryan}, {Tsangarides}, {Norris},
  {Beers} \& {Ando}}{{Aoki} et~al.}{2003}]{2003IAUJD..15E..19A}
{Aoki} W.,  {Ryan} S.~G.,  {Tsangarides} S.,  {Norris} J.~E.,  {Beers} T.~C.,
   {Ando} H.,  2003, Elemental Abundances in Old Stars and Damped
  Lyman-{$\alpha$} Systems, 25th meeting of the IAU, Joint Discussion 15, 22
  July 2003, Sydney, Australia, 15

\bibitem[\protect\citeauthoryear{{Arnould}, {Goriely} \& {Jorissen}}{{Arnould}
  et~al.}{1999}]{1999A&A...347..572A}
{Arnould} M.,  {Goriely} S.,    {Jorissen} A.,  1999, A\&A, 347, 572

\bibitem[\protect\citeauthoryear{{Beers} \& {Christlieb}}{{Beers} \&
  {Christlieb}}{2005}]{2005ARA&A..43..531B}
{Beers} T.~C.,  {Christlieb} N.,  2005, ARA\&A, 43, 531

\bibitem[\protect\citeauthoryear{{Campbell} \& {Lattanzio}}{{Campbell} \&
  {Lattanzio}}{2008}]{2008A&A...490..769C}
{Campbell} S.~W.,  {Lattanzio} J.~C.,  2008, A\&A, 490, 769

\bibitem[\protect\citeauthoryear{{Cayrel}, {Depagne}, {Spite}, {Hill}, {Spite},
  {Fran{\c c}ois}, {Plez}, {Beers}, {Primas}, {Andersen}, {Barbuy},
  {Bonifacio}, {Molaro} \& {Nordstr{\"o}m}}{{Cayrel}
  et~al.}{2004}]{2004A&A...416.1117C}
{Cayrel} R.,  {Depagne} E.,  {Spite} M.,  {Hill} V.,  {Spite} F.,  {Fran{\c
  c}ois} P.,  {Plez} B.,  {Beers} T.,  {Primas} F.,  {Andersen} J.,  {Barbuy}
  B.,  {Bonifacio} P.,  {Molaro} P.,    {Nordstr{\"o}m} B.,  2004, A\&A, 416,
  1117

\bibitem[\protect\citeauthoryear{{Charbonnel} \& {Zahn}}{{Charbonnel} \&
  {Zahn}}{2007}]{2007A&A...467L..15C}
{Charbonnel} C.,  {Zahn} J.-P.,  2007, A\&A, 467, L15

\bibitem[\protect\citeauthoryear{{Cohen}, {Christlieb}, {McWilliam},
  {Shectman}, {Thompson}, {Wasserburg}, {Ivans}, {Dehn}, {Karlsson} \&
  {Melendez}}{{Cohen} et~al.}{2004}]{2004ApJ...612.1107C}
{Cohen} J.~G.,  {Christlieb} N.,  {McWilliam} A.,  {Shectman} S.,  {Thompson}
  I.,  {Wasserburg} G.~J.,  {Ivans} I.,  {Dehn} M.,  {Karlsson} T.,
  {Melendez} J.,  2004, ApJ, 612, 1107

\bibitem[\protect\citeauthoryear{{Denissenkov} \& {Pinsonneault}}{{Denissenkov}
  \& {Pinsonneault}}{2008a}]{2008ApJ...684..626D}
{Denissenkov} P.~A.,  {Pinsonneault} M.,  2008a, ApJ, 684, 626

\bibitem[\protect\citeauthoryear{{Denissenkov} \& {Pinsonneault}}{{Denissenkov}
  \& {Pinsonneault}}{2008b}]{2008ApJ...679.1541D}
{Denissenkov} P.~A.,  {Pinsonneault} M.,  2008b, ApJ, 679, 1541

\bibitem[\protect\citeauthoryear{{Eggleton}}{{Eggleton}}{1971}]{1971MNRAS.151.%
.351E}
{Eggleton} P.~P.,  1971, MNRAS, 151, 351

\bibitem[\protect\citeauthoryear{{Eggleton}, {Dearborn} \&
  {Lattanzio}}{{Eggleton} et~al.}{2006}]{2006Sci...314.1580E}
{Eggleton} P.~P.,  {Dearborn} D.~S.~P.,    {Lattanzio} J.~C.,  2006, Science,
  314, 1580

\bibitem[\protect\citeauthoryear{{Eldridge} \& {Tout}}{{Eldridge} \&
  {Tout}}{2004}]{2004MNRAS.348..201E}
{Eldridge} J.~J.,  {Tout} C.~A.,  2004, MNRAS, 348, 201

\bibitem[\protect\citeauthoryear{{Frost} \& {Lattanzio}}{{Frost} \&
  {Lattanzio}}{1996}]{1996ApJ...473..383F}
{Frost} C.~A.,  {Lattanzio} J.~C.,  1996, ApJ, 473, 383

\bibitem[\protect\citeauthoryear{{Fujimoto}, {Iben} \& {Hollowell}}{{Fujimoto}
  et~al.}{1990}]{1990ApJ...349..580F}
{Fujimoto} M.~Y.,  {Iben} I.~J.,    {Hollowell} D.,  1990, ApJ, 349, 580

\bibitem[\protect\citeauthoryear{{Goswami}}{{Goswami}}{2005}]{2005MNRAS.359..5%
31G}
{Goswami} A.,  2005, MNRAS, 359, 531

\bibitem[\protect\citeauthoryear{{Herwig}}{{Herwig}}{2000}]{2000A&A...360..952%
H}
{Herwig} F.,  2000, A\&A, 360, 952

\bibitem[\protect\citeauthoryear{{Herwig}}{{Herwig}}{2004}]{2004ApJS..155..651%
H}
{Herwig} F.,  2004, ApJS., 155, 651

\bibitem[\protect\citeauthoryear{{Iben} Jr. \& {Renzini}}{{Iben} \&
  {Renzini}}{1983}]{1983ARA&A..21..271I}
{Iben} Jr. I.,  {Renzini} A.,  1983, ARAA, 21, 271

\bibitem[\protect\citeauthoryear{{Iglesias} \& {Rogers}}{{Iglesias} \&
  {Rogers}}{1996}]{1996ApJ...464..943I}
{Iglesias} C.~A.,  {Rogers} F.~J.,  1996, ApJ, 464, 943

\bibitem[\protect\citeauthoryear{{Johnson}, {Herwig}, {Beers} \&
  {Christlieb}}{{Johnson} et~al.}{2007}]{2007ApJ...658.1203J}
{Johnson} J.~A.,  {Herwig} F.,  {Beers} T.~C.,    {Christlieb} N.,  2007, ApJ,
  658, 1203

\bibitem[\protect\citeauthoryear{{Karakas} \& {Lattanzio}}{{Karakas} \&
  {Lattanzio}}{2007}]{2007PASA...24..103K}
{Karakas} A.,  {Lattanzio} J.~C.,  2007, Publications of the Astronomical
  Society of Australia, 24, 103

\bibitem[\protect\citeauthoryear{{Karakas}, {Lee}, {Lugaro}, {G{\"o}rres} \&
  {Wiescher}}{{Karakas} et~al.}{2008}]{2008ApJ...676.1254K}
{Karakas} A.~I.,  {Lee} H.~Y.,  {Lugaro} M.,  {G{\"o}rres} J.,    {Wiescher}
  M.,  2008, ApJ, 676, 1254

\bibitem[\protect\citeauthoryear{{Kippenhahn}, {Ruschenplatt} \&
  {Thomas}}{{Kippenhahn} et~al.}{1980}]{1980A&A....91..175K}
{Kippenhahn} R.,  {Ruschenplatt} G.,    {Thomas} H.-C.,  1980, A\&A, 91, 175

\bibitem[\protect\citeauthoryear{{Korn}, {Grundahl}, {Richard}, {Mashonkina},
  {Barklem}, {Collet}, {Gustafsson} \& {Piskunov}}{{Korn}
  et~al.}{2007}]{2007ApJ...671..402K}
{Korn} A.~J.,  {Grundahl} F.,  {Richard} O.,  {Mashonkina} L.,  {Barklem}
  P.~S.,  {Collet} R.,  {Gustafsson} B.,    {Piskunov} N.,  2007, ApJ, 671, 402

\bibitem[\protect\citeauthoryear{{Lind}, {Korn}, {Barklem} \&
  {Grundahl}}{{Lind} et~al.}{2008}]{2008A&A...490..777L}
{Lind} K.,  {Korn} A.~J.,  {Barklem} P.~S.,    {Grundahl} F.,  2008, A\&A, 490,
  777

\bibitem[\protect\citeauthoryear{{Lucatello}, {Beers}, {Christlieb}, {Barklem},
  {Rossi}, {Marsteller}, {Sivarani} \& {Lee}}{{Lucatello}
  et~al.}{2006}]{2006ApJ...652L..37L}
{Lucatello} S.,  {Beers} T.~C.,  {Christlieb} N.,  {Barklem} P.~S.,  {Rossi}
  S.,  {Marsteller} B.,  {Sivarani} T.,    {Lee} Y.~S.,  2006, ApJ.Lett., 652,
  L37

\bibitem[\protect\citeauthoryear{{Lucatello}, {Gratton}, {Beers} \&
  {Carretta}}{{Lucatello} et~al.}{2005}]{2005ApJ...625..833L}
{Lucatello} S.,  {Gratton} R.~G.,  {Beers} T.~C.,    {Carretta} E.,  2005, ApJ,
  625, 833

\bibitem[\protect\citeauthoryear{{Lucatello}, {Tsangarides}, {Beers},
  {Carretta}, {Gratton} \& {Ryan}}{{Lucatello}
  et~al.}{2005}]{2005ApJ...625..825L}
{Lucatello} S.,  {Tsangarides} S.,  {Beers} T.~C.,  {Carretta} E.,  {Gratton}
  R.~G.,    {Ryan} S.~G.,  2005, Ap.J., 625, 825

\bibitem[\protect\citeauthoryear{{Lugaro}, {de Mink}, {Izzard}, {Campbell},
  {Karakas}, {Cristallo}, {Pols}, {Lattanzio}, {Straniero}, {Gallino} \&
  {Beers}}{{Lugaro} et~al.}{2008}]{2008A&A...484L..27L}
{Lugaro} M.,  {de Mink} S.~E.,  {Izzard} R.~G.,  {Campbell} S.~W.,  {Karakas}
  A.~I.,  {Cristallo} S.,  {Pols} O.~R.,  {Lattanzio} J.~C.,  {Straniero} O.,
  {Gallino} R.,    {Beers} T.~C.,  2008, A\&A, 484, L27

\bibitem[\protect\citeauthoryear{{Lugaro}, {Ugalde}, {Karakas}, {G{\" o}rres},
  {Wiescher}, {Lattanzio} \& {Cannon}}{{Lugaro}
  et~al.}{2004}]{2004ApJ...615..934L}
{Lugaro} M.,  {Ugalde} C.,  {Karakas} A.~I.,  {G{\" o}rres} J.,  {Wiescher} M.,
   {Lattanzio} J.~C.,    {Cannon} R.~C.,  2004, ApJ, 615, 934

\bibitem[\protect\citeauthoryear{{Mel{\'e}ndez} \& {Cohen}}{{Mel{\'e}ndez} \&
  {Cohen}}{2007}]{2007ApJ...659L..25M}
{Mel{\'e}ndez} J.,  {Cohen} J.~G.,  2007, ApJ, 659, L25

\bibitem[\protect\citeauthoryear{{Meynet}, {Ekstr{\"o}m} \& {Maeder}}{{Meynet}
  et~al.}{2006}]{2006A&A...447..623M}
{Meynet} G.,  {Ekstr{\"o}m} S.,    {Maeder} A.,  2006, A\&A, 447, 623

\bibitem[\protect\citeauthoryear{{Pols}, {Tout}, {Eggleton} \& {Han}}{{Pols}
  et~al.}{1995}]{1995MNRAS.274..964P}
{Pols} O.~R.,  {Tout} C.~A.,  {Eggleton} P.~P.,    {Han} Z.,  1995, MNRAS, 274,
  964

\bibitem[\protect\citeauthoryear{{Richard}, {Michaud} \& {Richer}}{{Richard}
  et~al.}{2002}]{2002ApJ...580.1100R}
{Richard} O.,  {Michaud} G.,    {Richer} J.,  2002, ApJ, 580, 1100

\bibitem[\protect\citeauthoryear{{Richard}, {Michaud} \& {Richer}}{{Richard}
  et~al.}{2005}]{2005ApJ...619..538R}
{Richard} O.,  {Michaud} G.,    {Richer} J.,  2005, ApJ, 619, 538

\bibitem[\protect\citeauthoryear{{Roederer}, {Frebel}, {Shetrone}, {Allende
  Prieto}, {Rhee}, {Gallino}, {Bisterzo}, {Sneden}, {Beers} \&
  {Cowan}}{{Roederer} et~al.}{2008}]{2008ApJ...679.1549R}
{Roederer} I.~U.,  {Frebel} A.,  {Shetrone} M.~D.,  {Allende Prieto} C.,
  {Rhee} J.,  {Gallino} R.,  {Bisterzo} S.,  {Sneden} C.,  {Beers} T.~C.,
  {Cowan} J.~J.,  2008, ApJ, 679, 1549

\bibitem[\protect\citeauthoryear{{Schuler}, {Cunha}, {Smith}, {Sivarani},
  {Beers} \& {Lee}}{{Schuler} et~al.}{2007}]{2007ApJ...667L..81S}
{Schuler} S.~C.,  {Cunha} K.,  {Smith} V.~V.,  {Sivarani} T.,  {Beers} T.~C.,
   {Lee} Y.~S.,  2007, ApJ, 667, L81

\bibitem[\protect\citeauthoryear{{Spite} \& {Spite}}{{Spite} \&
  {Spite}}{1982}]{1982A&A...115..357S}
{Spite} F.,  {Spite} M.,  1982, A\&A, 115, 357

\bibitem[\protect\citeauthoryear{{Stancliffe}}{{Stancliffe}}{2005}]{stancliffe%
05}
{Stancliffe} R.~J.,  2005, PhD thesis, University of Cambridge

\bibitem[\protect\citeauthoryear{{Stancliffe}}{{Stancliffe}}{2006}]{2006MNRAS.%
370.1817S}
{Stancliffe} R.~J.,  2006, MNRAS, 370, 1817

\bibitem[\protect\citeauthoryear{{Stancliffe} \& {Glebbeek}}{{Stancliffe} \&
  {Glebbeek}}{2008}]{2008MNRAS.389.1828S}
{Stancliffe} R.~J.,  {Glebbeek} E.,  2008, MNRAS, 389, 1828

\bibitem[\protect\citeauthoryear{{Stancliffe}, {Glebbeek}, {Izzard} \&
  {Pols}}{{Stancliffe} et~al.}{2007}]{2007A&A...464L..57S}
{Stancliffe} R.~J.,  {Glebbeek} E.,  {Izzard} R.~G.,    {Pols} O.~R.,  2007,
  A\&A, 464, L57

\bibitem[\protect\citeauthoryear{{Stancliffe}, {Izzard} \& {Tout}}{{Stancliffe}
  et~al.}{2005}]{2005MNRAS.356L...1S}
{Stancliffe} R.~J.,  {Izzard} R.~G.,    {Tout} C.~A.,  2005, MNRAS, 356, L1

\bibitem[\protect\citeauthoryear{{Stancliffe} \& {Jeffery}}{{Stancliffe} \&
  {Jeffery}}{2007}]{2007MNRAS.375.1280S}
{Stancliffe} R.~J.,  {Jeffery} C.~S.,  2007, MNRAS, 375, 1280

\bibitem[\protect\citeauthoryear{{Stancliffe}, {Lugaro}, {Ugalde}, {Tout},
  {G\"orres} \& {Wiescher}}{{Stancliffe} et~al.}{2005}]{2005MNRAS.360..375S}
{Stancliffe} R.~J.,  {Lugaro} M.~A.,  {Ugalde} C.,  {Tout} C.~A.,  {G\"orres}
  J.,    {Wiescher} M.,  2005, MNRAS, 360, 375

\bibitem[\protect\citeauthoryear{{Stancliffe}, {Tout} \& {Pols}}{{Stancliffe}
  et~al.}{2004}]{2004MNRAS.352..984S}
{Stancliffe} R.~J.,  {Tout} C.~A.,    {Pols} O.~R.,  2004, MNRAS, 352, 984

\bibitem[\protect\citeauthoryear{{Suda}, {Katsuta}, {Yamada}, {Suwa},
  {Ishizuka}, {Komiya}, {Sorai}, {Aikawa} \& {Fujimoto}}{{Suda}
  et~al.}{2008}]{2008arXiv0806.3697S}
{Suda} T.,  {Katsuta} Y.,  {Yamada} S.,  {Suwa} T.,  {Ishizuka} C.,  {Komiya}
  Y.,  {Sorai} K.,  {Aikawa} M.,    {Fujimoto} M.~Y.,  2008, PASJ, 60, 1159

\bibitem[\protect\citeauthoryear{{Thompson}, {Ivans}, {Bisterzo}, {Sneden},
  {Gallino}, {Vauclair}, {Burley}, {Shectman} \& {Preston}}{{Thompson}
  et~al.}{2008}]{2008ApJ...677..556T}
{Thompson} I.~B.,  {Ivans} I.~I.,  {Bisterzo} S.,  {Sneden} C.,  {Gallino} R.,
  {Vauclair} S.,  {Burley} G.~S.,  {Shectman} S.~A.,    {Preston} G.~W.,  2008,
  ApJ, 677, 556

\bibitem[\protect\citeauthoryear{{Ulrich}}{{Ulrich}}{1972}]{1972ApJ...172..165%
U}
{Ulrich} R.~K.,  1972, ApJ, 172, 165

\bibitem[\protect\citeauthoryear{{Umeda} \& {Nomoto}}{{Umeda} \&
  {Nomoto}}{2005}]{2005ApJ...619..427U}
{Umeda} H.,  {Nomoto} K.,  2005, ApJ, 619, 427

\bibitem[\protect\citeauthoryear{{Uttenthaler}, {Lebzelter}, {Palmerini},
  {Busso}, {Aringer} \& {Lederer}}{{Uttenthaler}
  et~al.}{2007}]{2007A&A...471L..41U}
{Uttenthaler} S.,  {Lebzelter} T.,  {Palmerini} S.,  {Busso} M.,  {Aringer} B.,
     {Lederer} M.~T.,  2007, A\&A, 471, L41

\bibitem[\protect\citeauthoryear{{Yong}, {Lambert} \& {Ivans}}{{Yong}
  et~al.}{2003}]{2003ApJ...599.1357Y}
{Yong} D.,  {Lambert} D.~L.,    {Ivans} I.~I.,  2003, ApJ, 599, 1357

\end{thebibliography}

\label{lastpage}
\end{document}